\documentclass[11pt, a4paper]{article}
\usepackage{jheppub}
\usepackage{graphicx}
\usepackage{epsfig,amsmath,amsthm}

\newcommand{\be}{\begin{equation}}
\newcommand{\ee}{\end{equation}}
\newcommand{\bea}{\begin{eqnarray}}
\newcommand{\eea}{\end{eqnarray}}
\def\bse{\begin{subequations}}
\def\ese{\end{subequations}}

\def\IZ{\relax\ifmmode\hbox{Z\kern-.4em Z}\else{Z\kern-.4em Z}\fi}

\newcommand{\non}{\nonumber \\}

\def\half{\frac{1}{2}} 

\def\del{{\partial}}

\newcommand{\h}[1]{{\hat #1}} 

\def\hA{\hat{A}}

\def\hphi{{\hat \phi}} \def\hgamma{{\hat \gamma}}

\def\td{\tilde{d}}

\def\tgamma{\widetilde{\gamma}}
\def\al{\alpha} \def\bt{\beta}

\def\presub{\vspace{.5cm} \noindent}

\def\bi{\begin{itemize}} \def\ei{\end{itemize}}

\def\({\left(} \def\){\right)}
\def\[{\left[} \def\]{\right]}

\title{\center{Comparing space+time decompositions in the post-Newtonian limit}}

\author{Barak Kol$^a$, Michele Levi$^a$ and Michael Smolkin$^b$\\
$^a$ {\it Racah Institute of Physics, Hebrew University, Jerusalem 91904, Israel} {\tt barak\_kol,michele@phys.huji.ac.il} \\
$^b$ {\it Perimeter Institute for Theoretical Physics, Waterloo,
Ontario N2L 2Y5, Canada} {\tt msmolkin@perimeterinstitute.ca}
 }

\abstract{
The relationship between the Arnowitt-Deser-Misner (ADM) field
decomposition and the non-relativistic gravitational (NRG) fields
attracted considerable interest recently. This paper compares the
two, especially with respect to
 computing the two-body post-Newtonian (PN)
effective action within the effective field theory (EFT) approach.
Both are space+time decompositions and hence do better than using
the standard metric. However, ADM is essentially a reduction over
space whereas NRG is essentially a reduction over time. We use a
variant of ADM which is linearly equivalent to NRG and the two are
identical at order 1PN. We compare the two at order 2PN and find
that ADM requires the computation of an additional Feynman
diagram. We argue that the computational excess will further
increase at higher orders.}

\begin{document}

\maketitle

\section{Introduction}

The effective field theory (EFT) approach to the two-body
post-Newtonian (PN) dynamics in general relativity (GR) was put
forward in \cite{GoldbergerRothstein1}, borrowing ideas from
effective quantum field theories. It is based on a hierarchy of
scales in the problem and essentially replaces the traditional
method of dealing with finite-size objects through matched
asymptotic expansion by introducing instead effective interactions
of point particles with their background. See \cite{DamourFarese}
for early precursors of the EFT approach to GR. The ability of the
method to go beyond the state of the art was
 demonstrated in \cite{PortoRothstein} \footnote{Even if
imperfectly, since it was missing certain contributions found in
\cite{SteinhoffHergtSchaefer} using Hamiltonian methods and also
found later in \cite{Porto:2008tb} to arise from indirect
contributions in the EFT method. See \cite{Levi:2008nh} for a
derivation using NRG fields (to be introduced below).} by
computing for the first time the next-to-leading spin(1)-spin(2)
interaction in the effective two-body action.

In the post-Newtonian limit, spacetime is nearly flat and
velocities are slow; hence, there is a distinguished time direction
and a space+time decomposition is useful. For that purpose
non-relativistic gravitational (NRG) fields were introduced in
\cite{CLEFT-caged,NRG} \footnote{See \cite{NRGprecursors} for
early precursors of the NRG fields.}
 and were used to give what is
probably the shortest derivation of the leading post-Newtonian
correction, known as the Einstein-Infeld-Hoffmann interaction
\cite{EIH}. Recently, the full Einstein-Hilbert action was obtained
in terms of these fields \cite{NRGaction}. Related interesting and
relatively recent work on PN and/or EFT appeared in
\cite{related}.

The definition of NRG fields has some similarities with the
well-known Arnowitt-Deser-Misner (ADM) fields \cite{ADM}. Actually
the two definitions are identical in the static limit (where in
particular $g_{0i}=0$) and more generally they are linearly
equivalent (at least after some modification of ADM as we review
below) and therefore the two-body effective action is identical up
to order 1PN.
This partial similarity attracted considerable interest and raised
the question whether the distinction is
essential.\footnote{Private communications.} The purpose of this
paper is to explain the differences both from a conceptual point
of view and from a practical and computational PN
perspective.

While both NRG and ADM represent space+time decomposition, they are
nonlinearly inequivalent and there is a marked difference: NRG is
essentially a temporal Kaluza-Klein \cite{Kaluza-Klein} reduction
while ADM is essentially a spatial reduction.
Since the PN spacetime is nearly stationary, a temporal reduction
is conceptually fit. Yet, from a practical point of view it is
desirable to compare the two field definitions as they function
while computing the effective two-body PN action. We use the
harmonic gauge to enable comparison with previous derivations. Due
to the 1PN equivalence, we proceed to order 2PN (see the review
\cite{BlanchetRev} and references therein). The 2PN action was
reproduced within the EFT approach in \cite{GilmoreRoss} who found
the NRG fields to be preferable for their computations.

In this paper, we reproduce the 2PN effective action using the ADM
fields and compare with \cite{GilmoreRoss}. In section
\ref{sec:def-n-action}, we present the field definitions and
action, in section \ref{sec:eftADM} the Feynman rules, and in
section \ref{sec:2PN} we evaluate the required Feynman diagrams.
Finally our summary and discussions are presented in section
\ref{sec:summary}.

\section{Field definition and action}
 \label{sec:def-n-action}

\noindent{\bf Field definition}. The non-relativistic
gravitational (NRG) fields were introduced in
\cite{CLEFT-caged,NRG} through a temporal Kaluza-Klein
\cite{Kaluza-Klein} reduction followed by a Weyl rescaling \be
ds^2 = e^{2 \hphi}(dt - \hA_i\, dx^i)^2 -e^{-2 \hphi/\td}\,
\hgamma_{ij}\, dx^i dx^j ~, \label{def-NRG}  \ee where for greater
generality we work with an arbitrary spacetime dimension and we
denote \be \td:= d-3 ~. \ee Here the NRG fields are hatted to
distinguish them from the ADM fields below. The temporal reduction
divides the metric into a scalar, a vector and a tensor with
respect to \emph{spatial} transformations, while time shifts gauge
the vector. The Weyl rescaling leading to $\hgamma_{ij}$ is
performed to decouple the quadratic action and hence the
associated propagators (equivalently, $\hgamma_{ij}$ appears in
the action in a canonical Einstein-Hilbert form).

The utility of the NRG fields received strong support from \cite{GilmoreRoss} who
reproduced the 2PN effective action through the EFT approach
choosing to work with NRG fields. They found essentially two
advantages for NRG fields over the standard metric: the quadratic
decoupling (this is especially useful since at leading order a
compact object couples only to $\hphi$), and the elimination of
certain bulk vertices.

The standard ADM decomposition \cite{ADM} (see also the review
\cite{3+1rev} and references therein) is given by  \be
 ds^2 = \al^2 dt^2 -  \widetilde{\gamma}_{ij}\, (dx^i +\bt^i dt) (dx^j +\bt^j dt) ~, \label{def-ADM}  \ee
 where $\al$ is the lapse and $\bt^i$ is the shift vector. It is
designed for the initial value problem
or time evolution.
Actually one notices that in ADM, the shift is on the spatial
coordinates, while in standard KK it is on the reduced coordinate,
for instance, time in NRG (\ref{def-NRG}). In this sense, ADM is
nothing but a (Kaluza-Klein) reduction over the spatial
directions. From this perspective of reducing over space and
concentrating on the time dynamics, the ADM fields have the
following transformation properties: $\widetilde{\gamma}_{ij}$ is
a matrix of temporal scalars, the shift $\bt^i$ is a set of
temporal vectors $\sim (h_t)^i $, and $\al^2$ is the temporal
metric $\sim h_{tt}$. Being 1D, the (temporal) vectors and tensor
($\bt^i$ and $\al^2$) are non-dynamic during time evolution. The
last statement can be both familiar and surprising from the point
of view of numerical relativity. Familiar because given
geometrical and source-free initial conditions on a Cauchy surface
the lapse and shift are indeed non-dynamic and can be chosen at
will (gauge choice). On the other hand, it could be surprising
because in the presence of material bodies, the fields including
the lapse and shift are determined (in a specified gauge, say
harmonic) by elliptic equations with material sources and are not
free to choose. The resolution of this tension is to remember that
in the latter case, the dynamic degrees of freedom are actually
with the material bodies and the gravitational field merely reacts
to it, at least in the near zone where radiation is unimportant.


Standard ADM has two clear though fixable drawbacks from the PN
perspective. These are a non-flat kinetic metric for $\al$ (namely
the $\al$ dependent prefactor in the kinetic term $S \supset \int
\al^{-2} |\del \al|^2$), and a mix between $\al$ and
$\widetilde{\gamma}:={\rm tr}(\widetilde{\gamma}_{ij})$ at the
quadratic level of the form $S \supset \del \al\, \del
\widetilde{\gamma}$. Both issues produce extra diagrams already at
order 1PN and the problem only aggravates with increasing PN
order: the $\al$ kinetic term produces a triple vertex for $\al$
which contributes starting with a Y-shaped diagram at 1PN, while
the mix adds a 2-vertex which appears at 1PN through the diagram
which describes an exchange of $\al$ which transforms into
$\widetilde{\gamma}$ and back into $\al$. However, these drawbacks
can be fixed by a ``modified ADM decomposition'' which we proceed
to define, in order to make a more essential and elaborate
comparison with the NRG decomposition.

We define
a modified ADM decomposition as follows \be
 ds^2 = e^{2 \phi}dt^2 - e^{-2 \phi/\td}
\gamma_{ij}(dx^i+A^idt) (dx^j+A^jdt)  \label{def-ADMmod} ~. \ee
This is obtained from the standard definition in equation (\ref{def-ADM}) by
defining $\phi:= \log (\al)$ in order to have a flat kinetic
metric for the scalar (namely $S \supset \int |\del \phi|^2$
rather than $S \supset \int \al^{-2} |\del \al|^2$); a Weyl
rescaled $\gamma_{ij}:=\exp(2\phi/\td) \widetilde{\gamma}_{ij}$
as before to achieve decoupling in the quadratic action;; and
finally a change of notation $A^i:=\bt^i$ to make the ADM and NRG
notations similar.

Comparing the definitions of the modified ADM from equation (\ref{def-ADMmod})
and the NRG fields in equation (\ref{def-NRG}), we observe that they are similar in
two ways. First, for $\hA_i=A^i=0$ they are identical. Second,
linearizing around flat space we define \be
 \sigma_{ij}=\gamma_{ij}-\delta_{ij}~,  \label{def-sigma} \ee
and similarly for NRG fields
$\hat{\sigma}_{ij}=\hgamma_{ij}-\delta_{ij}$, and then the two
sets are linearly equivalent around flat space, namely
$\hphi=\phi+\dots, ~\hA_i=A_i + \dots, ~
\hat{\sigma}_{ij}=\sigma_{ij} + \dots$ where the ellipsis denote
terms which are quadratic or higher in the perturbation fields
$\phi,\, A_i,\, \sigma_{ij}$ (the spatial indices $i,j,\dots$ are
raised and lowered here with $\delta_{ij})$.


\presub {\bf Action}. The total action is a sum of three parts \be
 S_{tot}=S_{EH} + S_{GF} + S_p, \label{action-components} \ee
 where $S_{EH}$ is the bulk Einstein-Hilbert action, $S_{GF}$ is
the harmonic gauge-fixing term and $S_p$ is the particle action.

In order to obtain the Einstein-Hilbert action $S=-1/(16 \pi G)
\int R dV$ we first computed it for the original ADM form
of equation (\ref{def-ADM}) and then performed a field redefinition. For the
first step, we used the non-orthonormal frame method (see for
example \cite{NRGaction}) and found the well-known result \be
 S_{EH} = -\frac{1}{16 \pi G} \int \al \, \sqrt{\tgamma}\, d^{d-1}x\, dt
 \left\{  -  \( \left| K_{ij} \right|^2  - K^2\) - R[\tgamma]
 \right\}, \label{ADM-pre-W-action}
\ee
 where the extrinsic curvature is defined by \be
 K_{ij} = -\half \al^{-1} \( D_t \tgamma_{ij} - \tgamma_{k(j} \del_{i)} A^k \) \equiv -\half \al^{-1} \( \dot{\tgamma}_{ij} - {\cal L}_{A}
 \tgamma_{ij}\) ~,\ee
and where \be D_t := \del_t - A^i \del_i, \ee
and $\cal L_{A}$ denotes the Lie
derivative with respect to $A^i$. In addition, we define \bea
  K &:=& \tgamma^{ij} K_{ij}, \non
 \left| K_{ij} \right|^2 &:=&  K_{ij} K_{kl} \tgamma^{ik} \tgamma^{jl} ~.
\eea

In terms of the modified ADM fields in equation (\ref{def-ADMmod}), one finds
(note that this time the new $\gamma_{ij}$ appears, not the
pre-Weyl $\tgamma_{ij}$) \bea
 S_{EH} &=& -\frac{1}{16 \pi G} \int \sqrt{\gamma}\, d^{d-1}x\, dt
 \non
 && \left\{ \(1+\frac{1}{\td}\) \left| \del_i  \phi \right|^2  - R[\gamma]  \right.
 -e^{-2\phi/\td} \( \left| K_{ij}[\gamma] \right|^2  - K^2[\gamma]\)  \non &&
 + \left.  \left(1+ \frac{1}{\td}\right)\, e^{-2(1+1/\td)\phi}\, D_t \phi\( 2 e^\phi\, K[\gamma] + \left(1+ \frac{2}{\td}\right) D_t \phi \)  \right\}.
 \label{ADMmod-action} \eea
The harmonic gauge-fixing term is defined by \be
 S_{GF}= \frac{1}{2 \cdot 16 \pi G} \int e^{-2\phi/\td} \sqrt{\gamma}\, d^{d-1}x\, dt ~~  g^{ab}\, \Gamma_a\, \Gamma_b, \label{harmonic-gauge1} \ee
where $g^{ab} \Gamma_a \Gamma_b =\Gamma_{\h0}^2 - e^{2\phi/\td} \gamma^{ij} \Gamma_{\hat{i}} \Gamma_{\hat{j}}$ and $\h0,\hat{i}$ are frame indices.
In ADM variables, we find \bea
 \Gamma_{\h0} &=&  e^{-\phi}\, \del_i A^i - \half e^{-\phi}\, D_t \log \gamma + 2\(1+\frac{1}{\td}\) e^{-\phi}\, D_t \phi,   \non
 \Gamma_{\hat{i}} &=&  \Gamma_i[\gamma] - e^{-2(1+1/\td)\phi}\, \gamma_{ij}\, D_t
 A^j, \label{harmonic-gauge2}
\eea where $\Gamma_i[\gamma]=\gamma^{jk} \(\del_j \gamma_{ik} -
\del_i \gamma_{jk}/2 \)$, namely it is the contraction of the
Christoffel symbols for the metric $\gamma_{ij}$ which would be
used to define the standard harmonic gauge for $\gamma_{ij}$.

The worldline action is approximated at leading EFT order by a point
particle action and can be read from equation (\ref{def-ADMmod}) to be \be
 S_p \equiv -m \int d\tau = -m \int dt \sqrt{
 e^{2\phi}-e^{-2\phi/\td} \gamma_{ij} (v^i+A^i) (v^j+A^j)},
\label{particle-action}
\ee
where we denote the velocity by $v^i := dx^i/dt$.

\section{Feynman rules}
 \label{sec:eftADM}

The total action for the modified ADM fields of equation (\ref{def-ADMmod}) is
given by equations (\ref{action-components}) and
(\ref{ADMmod-action}-\ref{particle-action}) and we work in 4D for
definiteness. From it, we read the Feynman rules. The propagators
are
\begin{align}
\label{eq:prphi} \parbox{18mm}{\includegraphics{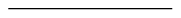}}
 & = \langle{~\phi(x_1)}~~{\phi(x_2)~}\rangle =~~~ 4 \pi\, G ~~~~ \int_{\bf{k}} \frac{e^{i{\bf k}\cdot\left({\bf x}_1 - {\bf x}_2\right)}}{{\bf k}^2}~\delta(t_1-t_2),\\
\label{eq:prA} \parbox{18mm}{\includegraphics{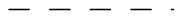}}
 & = \langle{A_i(x_1)}~{A_j(x_2)}\rangle = -  16 \pi\, G ~\delta_{ij} \int_{\bf{k}} \frac{e^{i{\bf k}\cdot\left({\bf x}_1 - {\bf x}_2\right)}}{{\bf k}^2}~\delta(t_1-t_2),\\
\label{eq:prsigma} \parbox{18mm}{\includegraphics{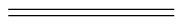}}
 & = \langle{\sigma_{ij}(x_1)}{\sigma_{kl}(x_2)}\rangle = 32 \pi\, G ~ P_{ij;kl}~ \int_{\bf{k}}\frac{e^{i{\bf k}\cdot\left({\bf x}_1 - {\bf x}_2\right)}}{{\bf k}^2}~\delta(t_1-t_2),
\end{align}
where $\int_{\bf{k}} \equiv \int \frac{d^3{\bf{k}}}{(2\pi)^3}$ for
abbreviation, and
$P_{ij;kl}\equiv\frac{1}{2}\left(\delta_{ik}\delta_{jl}+\delta_{il}\delta_{jk}-2\delta_{ij}\delta_{kl}\right)$.
Here and henceforth, the Feynman rules are presented in position
space. Note the simple form of the propagators obtained through
the exponentiation and Weyl rescaling in the modified definition
in equation (\ref{def-ADMmod}), especially for the $\phi$ and $A_i$ fields,
which dominate in the gravitational interaction. Note also that
factors of $16 \pi G$ could have been eliminated from the bulk
Feynman rules either by working with units such that $16 \pi G=1$
(or some other constant) or alternatively by computing $16 \pi G
~S_{eff}$ rather than $S_{eff}$.

In addition, there are time-dependent quadratic vertices. Mixed
vertices are eliminated by the Weyl rescaling together with the
gauge-fixing term. The Feynman rules for quadratic vertices
required up to order 2PN  are \begin{align} \label{eq:prtphi}
\parbox{18mm}{\includegraphics{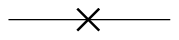}}
 & = ~~\frac{1}{8\pi G}~~\int d^4x~[\partial_t\phi(x)]^2, \\
\label{eq:prtA}   \parbox{18mm}{\includegraphics{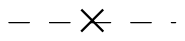}}
 & = -\frac{1}{32\pi G} \int d^4x~[\partial_tA_i(x)]^2.
\end{align}
The crosses represent bulk vertices that contain two time
derivatives. As expected, the Feynman rules up to quadratic order
are the same as those for the NRG fields.

The Feynman rules for the three-field bulk vertices required up to order 2PN  are \begin{align}
\label{eq:phiA^2} \parbox{18mm}{\includegraphics{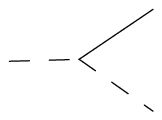}}
 & = -\frac{1}{8\pi G}\int d^4x~\biggl[\phi(x)\biggl(\partial_iA_j(x)(\partial_iA_j(x)+\partial_jA_i(x))-(\partial_iA_i(x))^2\biggr)\biggr], \\
\label{eq:sigmaphi^2} \parbox{18mm}{\includegraphics{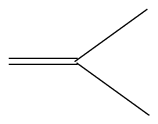}}
 & = ~ \frac{1}{16\pi G}\int  d^4x~[2\sigma_{ij}(x)\partial_i\phi(x)\partial_j\phi(x)-\sigma_{jj}(x)\partial_i\phi(x)\partial_i\phi(x)], \\
\label{eq:Aphi^2t} \parbox{18mm}{\includegraphics{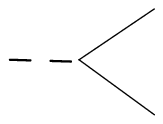}}
 & = -\frac{1}{4\pi G}\int d^4x~[A_i(x)\partial_i\phi(x)\partial_t\phi(x)],\\
\label{eq:phi^3dt^2} \parbox{18mm}{\includegraphics{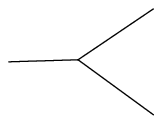}}
 & = -\frac{1}{2\pi G}\int d^4x~[\phi(x)(\partial_t\phi(x))^2],
\end{align}
where there is no distinction between lower and upper $i,j,\dots$
indices (or equivalently they are raised and lowered with
$\delta_{ij}$).

Note that only the $\phi A^2$ vertex in equation (\ref{eq:phiA^2}) is
\textit{different} than the respective vertex in terms of the NRG
fields. This is consistent with the fact that differences can only
occur in vertices which involve the field $A_i$. Yet the $A \,
\del \phi \, \dot{\phi}$ vertex turns out to be the same.

Now, we consider the gravitational coupling to the massive compact
objects. The Feynman rules for the one-field couplings to the
worldline mass required up to order 2PN are
\begin{align}
\label{eq:mphi} \parbox{12mm}{\includegraphics{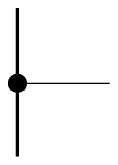}}
 & = - m \int dt~\phi({\bf{x}}(t))~\left[1+\frac{3}{2}v(t)^2+\frac{7}{8}v(t)^4+\cdots\right], \\
\label{eq:mA} \parbox{12mm}{\includegraphics{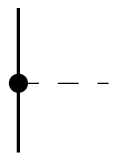}}
 & = ~m \int dt~A_i({\bf{x}}(t))v^i(t)~\left[1+\frac{1}{2}v(t)^2+\cdots\right], \\
\label{eq:msigma} \parbox{12mm}{\includegraphics{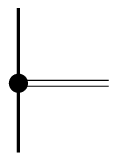}}
 & = ~\frac{m}{2} \int dt~\sigma_{ij}({\bf{x}}(t))v^i(t)v^j(t)~\left[1+\cdots\right],
\end{align}
where the heavy solid lines represent the worldlines, and the
spherical black vertices represent the masses on the worldline.
The ellipsis denotes higher orders in $v$, beyond the order
considered here.

The Feynman rules for two-field worldline vertices required up to order 2PN are
\begin{align}
\label{eq:mphi^2}  \parbox{12mm}{\includegraphics{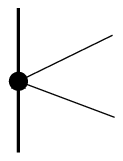}}
 & = -\frac{m}{2} \int dt~\phi({\bf{x}}(t))^2~\left[1-\frac{9}{2}v(t)^2+\cdots\right], \\
\label{eq:mA^2}  \parbox{12mm}{\includegraphics{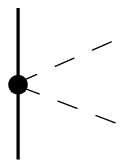}}
 & = \frac{m}{2} \int dt~A({\bf{x}}(t))^2~\left[1+\cdots\right], \\
\label{eq:mphiA}   \parbox{12mm}{\includegraphics{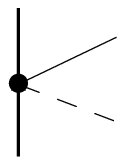}}
 & = -3m \int dt~\phi({\bf{x}}(t))A_i({\bf{x}}(t))v^i(t)~\biggl[1+\cdots\biggr].
\end{align}
Note the appearance of an $A^2$ two-field
worldline coupling in equation (\ref{eq:mA^2}) already at leading order in $v$ unlike the case for NRG fields, and that the $\phi A$ two-field worldline coupling
in equation (\ref{eq:mphiA}) is \textit{different} as well (larger by a factor of $(-3)$). We note again that the differences occur only in couplings which involve the field $A_i$.

Finally, the three-field couplings to the worldline required up to
order 2PN are represented by the following Feynman rule:
\begin{align}
\label{eq:mphi^3}  \parbox{12mm}{\includegraphics{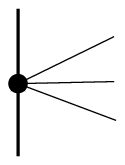}}
 & = -\frac{m}{6} \int dt~\phi({\bf{x}}(t))^3~\left[1+\cdots\right].
\end{align}

\section{Feynman diagrams and their evaluation at 2PN}
 \label{sec:2PN}

The differences in the EFT calculation of the binary interaction
start to appear at order 2PN (similarly, they appear already in
the next-to-leading order of the spin-orbit interaction). That is
so because all the Feynman rules required up to 1PN are identical:
most rules are identical due to linear equivalence whereas the
$\phi^2$ worldline coupling from equation (\ref{eq:mphi^2}) coincides due to
static equivalence (namely for $A^i=0$). The differences at 2PN
occur only in two topologies (out of the eight topologies contributing
at this order): the V topology of two-field exchange, and the Y
topology of the cubic vertices.

\begin{figure}[t]
\begin{center}
\includegraphics{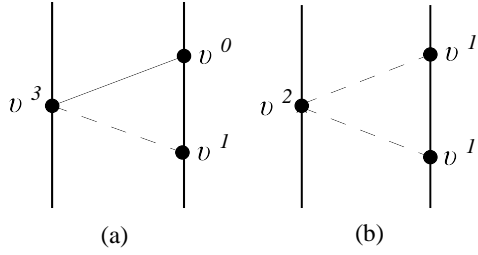}
\caption{2PN Feynman diagrams of two-field exchange including
differences from diagrams with NRG fields. (b) An additional
diagram that is eliminated when NRG fields are used. These
diagrams should be included together with their mirror
images.}\label{2pnadm2m}
\end{center}
\end{figure}

\presub {\bf Two-field exchange}. Since the $\phi A$ two-field
mass coupling in equation (\ref{eq:mphiA}) is different than the
respective mass coupling in terms of NRG fields, the respective
Feynman diagram, shown in figure 1(a) has a different value,
given by \be figure~1(a) =
-12\frac{G^2m_1m_2(m_1+m_2)}{r^2}~{\bf{v}}_1\cdot{\bf{v}}_2, \ee
where here and henceforth a prefactor of $\int dt$ is suppressed
and omitted from diagram values. Moreover, the `new' $A^2$
two-field mass coupling in equation (\ref{eq:mA^2}) gives rise to an
additional Feynman diagram depicted in figure 1(b), and evaluated
to be \be
figure~1(b) = 8\frac{G^2m_1m_2}{r^2}~(m_1{\bf{v}}_1^2+m_2{\bf{v}}_2^2). \ee

\begin{figure}[t]
\begin{center}
\includegraphics{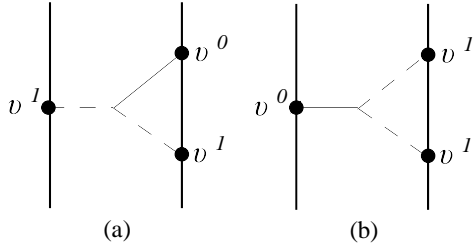}
\caption{2PN Feynman diagrams with a 3-field vertex including
differences from diagrams with NRG fields. These diagrams should
be included together with their mirror images.}\label{2pnadm3g}
\end{center}
\end{figure}

\presub {\bf Cubic gravitational interaction}. Since only the
$\phi A^2$ vertex in equation (\ref{eq:phiA^2}) differs from the
respective vertex in terms of NRG fields, the Feynman diagrams
including it, shown in figures 2(a), 2(b), have different values
given by \bea
figure~2(a) &=& 8\frac{G^2m_1m_2(m_1+m_2)}{r^2}~{\bf{v}}_1\cdot{\bf{v}}_2,\\
figure~2(b) &=& -4\frac{G^2m_1m_2}{r^2}~(m_1{\bf{v}}_1^2+m_2{\bf{v}}_2^2). \eea

\presub {\bf Total effective action at 2PN}. Altogether our EFT
calculation in terms of ADM-related fields reproduces the total
$S_{eff}$ at 2PN as found in \cite{GilmoreRoss} which used NRG
fields (in both cases, the harmonic gauge was used). The breakdown
of this net result is as follows. Only four diagrams are different
in the ADM calculation. Diagrams (h) and (k) in \cite{GilmoreRoss}
(see figure 5 and equations (31), (43) there) yield terms of the same
form, with coefficients\footnote{Note that \cite{GilmoreRoss}
compute the potential $V$ which is defined to be minus the value
of the diagram.} 4 and (-8), respectively, and a total of (-4).
This form appears here in figures 1(a) and 2(a).
Figure 1(a) yields the coefficient (-12) (recall that the vertex in
equation (\ref{eq:mphiA}) is rescaled by (-3) relative to \cite{GilmoreRoss}), and figure 2(a) yields\footnote{Recall the different vertex in equation (\ref{eq:phiA^2}).} the coefficient 8 with the same total of (-4). The diagrams in figures 1(b) and 2(b) also yield terms of the same form with coefficients 8 and (-4), respectively, and a total of 4, whereas in \cite{GilmoreRoss} the diagram corresponding to figure 1(b) here is not present, and figure 2(b) here is replaced (see footnote 7) by diagram (l) (see figure 5 and
equation (44) there) and is found to have a coefficient 4, which is identical with our total.

\section{Discussion}
 \label{sec:summary}

Comparing the computations of the 2PN effective action using the
modified ADM fields with the NRG fields, the Feynman rules are
equivalent up to quadratic order in the bulk vertices and linear
order in the worldline vertices. They differ for two vertices:
the $\phi\, A$ worldline vertex in equation (\ref{eq:mphiA}) and the bulk
$\phi\, A^2$ vertex in equation (\ref{eq:phiA^2}). In addition, there is a new
$A^2$ worldline vertex in equation (\ref{eq:mA^2}). Accordingly altogether, 18
of the diagrams are the same, 3 have different values but
comparable computational cost, and the ADM computation requires to
evaluate one extra diagram, namely that of figure 1(b), which is
factorizable (namely, the computation of the diagram factorizes
into a product, and each factor is associated with a sub-diagram).
It should be borne in mind that had we not modified the ADM fields
(along similar lines of the NRG fields definition), the
computational excess would have shown up already at 1PN.

\begin{figure}[t]
\begin{center}
\includegraphics{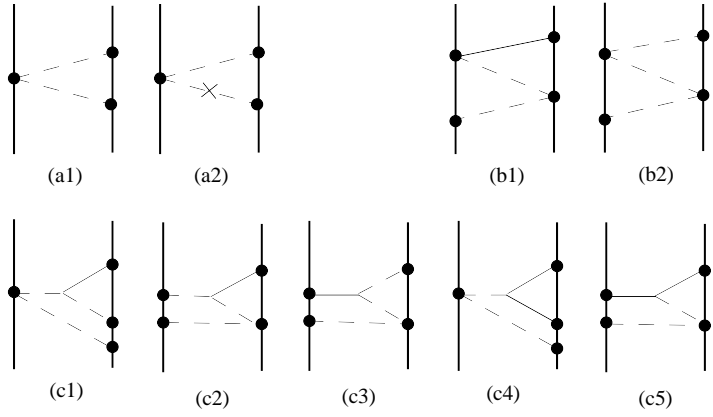}
\caption{Additional Feynman diagrams including the extra 2-field
$A^2$ worldline vertex, which appear at 3PN if the modified ADM
fields are used. In NRG fields, an $A^2$ vertex does not appear
altogether at 3PN, but rather it appears first at 4PN through the
term $(\vec{A} \cdot \vec{v})^2\, v^2$. Diagrams (a1)-(a2)
contribute at order $G^2$, while diagrams (b1)-(c5) contribute at
order $G^3$. Note that while diagram (a1) appears already at 2PN,
the velocity dependence of its vertices contributes also to 3PN.}
\label{3pnadm}
\end{center}
\end{figure}

 \presub {\bf At higher orders}. At higher orders,
the additional computational cost of the modified ADM fields will
further increase due to at least two reasons:
 first, the extra $A^2$ worldline vertex
from equation (\ref{eq:mA^2}) will require the computation of several
additional diagrams, as can be seen in figure 3.
In addition there are other worldline vertices
which appear at 3PN in ADM (but not in NRG), such as a new
$\sigma_{ij} A^i v^j$ worldline vertex.
Second, by comparing the ADM bulk action in
equations (\ref{ADMmod-action})-(\ref{harmonic-gauge2}) and the NRG
action in \cite{NRGaction} (incorporating the harmonic gauge), one finds that additional bulk vertices
will appear or get complicated at higher orders, and we give below
several examples.

The following vertices appear at 3PN and 4PN in ADM:
\begin{align}
\label{eq:nophiA^2}\parbox{18mm}{\includegraphics{frg6.eps}}
 & = -\frac{1}{8\pi G}\int d^4x~\left[\phi(x)\biggl(\partial_iA_j(x)(\partial_iA_j(x)+\partial_jA_i(x))-(\partial_iA_i(x))^2\biggr)\right. \non
 & \left. \,\,\,\,\,\,\,\,\,\,\,\,\,\,\,\,\,\,\,\,\,\,\,\,\,\,\,\,\,\,\,\,\,\,\,\,\,\,\,\,\,\,\,\,\,
-2\phi(x)(\partial_t A_i(x))^2\right],\\
\label{eq:phi^3A} \parbox{18mm}{\includegraphics{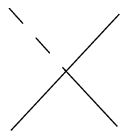}}
 & = \frac{1}{\pi G}\int d^4x~\left[\phi(x)\partial_t \phi(x) \partial_i \phi(x) A_i(x)\right],\\
\label{eq:A^4} \parbox{18mm}{\includegraphics{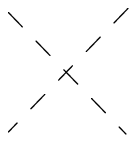}}
 & = -\frac{1}{32\pi G}\int d^4x~(A_i(x) \partial_i A_j(x))^2 ~.
\end{align}
In NRG, the first vertex is stationary and hence simpler, while the
quartic vertices are absent altogether.

\subsection*{Acknowledgments}

This research is supported by the Israel Science Foundation grant
no 607/05, by the German Israel Cooperation Project grant DIP
H.52, and by the Einstein Center at the Hebrew University. Research at
Perimeter Institute is supported by the Government of Canada through Industry Canada and
by the Province of Ontario through the Ministry of Research \& Innovation.

\end{document}